\documentclass[12pt,a4paper]{article}

\newcommand{\var}{\varphi}
\newcommand{\pa}{\partial}
\newcommand{\sn}{\mathrm{sn}}
\newcommand{\dn}{\mathrm{dn}}
\newcommand{\ka}{\kappa}
\newcommand{\cn}{\mathrm{cn}}

\begin{document}

\begin{flushright}
hep-th/0303237
\end{flushright}
\vspace{1.8cm}

\begin{center}
 \textbf{\Large Rotating and Orbiting Strings \\ 
in the Near-Horizon Brane Backgrounds}
\end{center}
\vspace{1.6cm}
\begin{center}
 Shijong Ryang
\end{center}

\begin{center}
\textit{Department of Physics \\ Kyoto Prefectural University of Medicine
\\ Taishogun, Kyoto 603-8334 Japan}  \par
\texttt{ryang@koto.kpu-m.ac.jp}
\end{center}
\vspace{2.8cm}
\begin{abstract}
Using the Schwarzschild-type coordinates in stead of the global ones we
reconstruct the classical rotating closed string solutions in the 
$AdS_5\times S^5$ backgrounds. They are explicitly described by the 
Jacobi elliptic and trigonometrical functions of worldsheet coordinates.
We study the orbiting closed string configurations in the near-horizon 
geometries of D$p$, NS1 and NS5 branes, and derive the energy and spin
of them, whose relation takes a simple form for short strings.
Specially in the D5 and NS5 backgrounds we have a linear relation that
the energy of the point-like string is proportional to the spin,
which is associated with the spectrum of strings in the pp-wave geometries
obtained by taking a special Penrose limit on the D5 and NS5 backgrounds.
\end{abstract} 
\vspace{3cm}
\begin{flushleft}
March, 2003
\end{flushleft}

\newpage
\section{Introduction}

The AdS/CFT correspondence relates the weakly coupled string theory 
on the $AdS_5\times S^5$ background to the strongly coupled large N
$\mathcal{N}=4$ super Yang-Mills theory \cite{JM,GKP,EW}.
This conjecture has been shown in the supergravity approximation
where the curvature is small and $\alpha'R$ can be neglected. 
However, in order to verify the correspondence in its full extent it is
desirable to go beyond the supergravity approximation. Based on the 
observation \cite{RM,MT} that the string theory in the pp-wave 
background \cite{BF} is solvable, it has been shown how to identify 
particular string states with gauge invariant operators with large 
R-charge from the gauge theory side \cite{BMN}.

Gubser, Klebanov and Polyakov \cite{SIA} have presented the semiclassical
approach to string motions in $AdS_5\times S^5$ in the global coordinates
and reproduced the results of \cite{BMN} by considering a particular 
configuration of classical rotating closed strings on $S^5$.
They have also studied a rotating string configuration on $AdS_5$
stretched along the radial direction to give an energy-spin relation for
this string state that can be compared with the perturbative result for
conformal gauge field theories. A general string motion rotating both on
$AdS_5$ and on $S^5$ in the global coordinates has been studied and a 
general formula relating the energy, spin and R-charge has been 
presented \cite{FT}, which has been 
further generalized to give interpolating
more general solutions \cite{JR}. Further interesting developments of the
semiclassical approach to strings, membranes and D-branes in the 
$AdS\times S$-type backgrounds can be found in
\cite{MSW,AG,JAM,AT,RV,LO,PO}, where
M- and string theories on this type of backgrounds are dual to 
conformal field theories on the brane worldvolumes.

The semiclassical approach is also expected to hold for more general cases
such as non- or less supersymmetric cases and non-conformal 
cases  \cite{ABP,AT,AJA,AM,HN,GZ,PT}.
The orbiting strings in a $AdS_5$ black hole background have been 
investigated and associated with the glueball states in the dual
finite temperature $\mathcal{N}$=4 SYM theory \cite{ABP}.
Rotating string configurations in a confining geometry produced by an 
AdS charged black hole have been also studied \cite{AJA}, 
where long strings probe the interplay between confinement and 
finite-size effect. Further in Witten's model for QCD constructed with
near-extremal D$p$-branes the string configurations rotating along a 
circle with a fixed radius in the non-compact directions of  
D$p$-branes, have been shown to give the dispersion relations between the
energy and the linear momentum along the compact circle \cite{AJA,AM}. 
The semiclassical approach has been applied to the membranes rotating  
in the $G_2$ holonomy backgrounds that are not $AdS\times S$-type ones
and dual to $\mathcal{N}$=1 gauge theories in four dimensions and 
various type of energy-spin relations have been presented \cite{HN},
while several folded closed string configurations have been studied
in the Maldacena-Nu\~nez background, dual in the infra-red to 
$\mathcal{N}$=1 gauge theories in four dimensions \cite{PT}.

Using the standard Schwarzschild-type coordinates in stead of 
the global ones we will reconstruct the classical 
solution representing a closed string rotating on $AdS_5$ in the 
$AdS_5\times S^5$ background. We will also consider the string 
soliton rotating on $S^5$
and stretched along an angular direction in the $AdS_5\times S^5$ 
background expressed by the Schwarzschild-type coordinates. This gives
an insight how to study the closed string configurations rotating along an
angular direction of the transverse space in the near-horizon geometries 
of D$p$-branes which are dual to the non-conformal field theories.
Further we will analyze the closed string motions in the near-horizon 
geometries of NS5-branes as well as NS1-branes and derive the 
energy-spin relations for the  various backgrounds.
Specially the energy-spin relations for point-like strings in the D5-brane
and NS5-brane backgrounds will be associated with the string spectra in
the pp-wave geometries obtained by taking a special 
Penrose limit on the D5-brane and NS5-brane backgrounds.

\section{Rotating strings in $AdS_5\times S^5$}

We consider a rotating closed string motion in the $AdS_5\times S^5$ 
background. In stead of the global metric we use the following metric in
the Schwarzschild-type coordinates
\begin{equation}
ds^2 = \frac{r^2}{R^2} ( -dt^2 + \sum_{i=1}^{3}dx_i^2 ) + 
\frac{R^2}{r^2}dr^2 + R^2(\sin^2\theta d\var^2 + d\theta^2 
+ \cos^2\theta d\Omega_3^2 ),
\end{equation}
where $R^2 = \sqrt{\lambda}\alpha'$ with the 't Hooft coupling $\lambda =
g_{YM}^2N$ and
\begin{eqnarray}
\sum_{i=1}^{3}dx_i^2 &=& dl^2 + l^2( d\phi_1 + \sin^2\phi_1 d\phi_2^2 ),
\nonumber \\ 
d\Omega_3^2 &=& d\var_1^2 + \cos^2\var_1 (d\var_2^2 + \cos^2\var_2
d\var_3^2 ).
\end{eqnarray}
In order to look for a classical string solution with conserved energy 
and angular momentum, we assume that a closed string rotates along 
$\phi = \phi_2$ of $AdS_5$ and make an ansatz for a time-dependent 
embedding
\begin{eqnarray}
t &=& t(\tau),\; l = l(\tau,\sigma),\; r = r(\tau,\sigma), \nonumber \\
\phi &=& \omega\tau, \;\phi_1 = \theta = \frac{\pi}{2}, \;\var_i = 0 \;
( i=1,2,3)
\label{tlr}\end{eqnarray}
satisfying periodicity conditions, $l(\sigma + 2\pi) = l(\sigma)$
and $r(\sigma + 2\pi) = r(\sigma)$, where $\tau$ and $\sigma$ denote
the worldsheet time and space coordinates, and the angular velocity 
parameter $\omega$ is constant. The relevant worldsheet action for the 
closed string embedding of the form (\ref{tlr}) is given by
\begin{eqnarray}
I &=& \int d\tau d\sigma L, \nonumber \\
L &=& - \frac{1}{4\pi\alpha'} \left[ \frac{r^2}{R^2}
(\dot{t}^2 - l^2\dot{\phi}^2 - \dot{l}^2 + l'^2 ) + 
\frac{R^2}{r^2} ( -\dot{r}^2 + r'^2 ) \right],
\label{ac}\end{eqnarray}
where dot and prime represent derivatives with respect to $\tau$ 
and $\sigma$ respectively.

The variations with respect to $t$ and $\phi$ lead, after one integration,
to the first order differential equations
\begin{eqnarray}
r^2\dot{t} &=& A, \label{ar} \\
r^2l^2\dot{\phi} &=& B, 
\label{br}\end{eqnarray}
where the integration constants $A, B$ are $\tau$-independent.
The string equations of motion for $l$ and $r$ read
\begin{eqnarray}
&\frac{\pa}{\pa\tau}(r^2\dot{l}) - \frac{\pa}{\pa\sigma}(r^2l')
- \dot{\phi}^2lr^2 = 0, \label{sts} \\
&R^2\left[ \frac{\pa}{\pa\tau} \left(\frac{\dot{r}}{r^2}\right)
- \frac{\pa}{\pa\sigma}\left(\frac{r'}{r^2}\right) \right] +
\frac{R^2}{r^3}(\dot{r}^2 - r'^2) + \frac{r}{R^2}( \dot{t}^2
- l^2\dot{\phi}^2 - \dot{l}^2 + l'^2 ) = 0,
\label{ste}\end{eqnarray}
while the constraints on the energy-momentum tensor of the system,
which are expressed by $T_{\tau\sigma} = 0$ and $T_{\tau\tau} + 
T_{\sigma\sigma} = 0$, provide
\begin{eqnarray}
&\frac{R^2}{r^2}\dot{r}r' + \frac{r^2}{R^2}\dot{l}l' = 0,
\label{ts} \\
&\frac{R^2}{r^2}(\dot{r}^2 + r'^2) + \frac{r^2}{R^2}( -\dot{t}^2
+ \omega^2l^2 + \dot{l}^2 + l'^2 ) = 0.
\label{tt}\end{eqnarray}
Substituting the separable forms as $r = f(\tau)g(\sigma)$ and $l = 
F(\tau)G(\sigma)$ into (\ref{ts}) we have
\begin{equation}
- R^4 \frac{\dot{f}}{F\dot{F}f^3} = \frac{GG'g^3}{g'},
\label{rfg}\end{equation}
which should be constant to be equated with $C_1$. From the requirement of
(\ref{br}) we get $F = 1/f$, which is substituted into (\ref{rfg}) 
to yield $C_1 = R^4$ together with
\begin{equation}
GG' = \frac{R^4g'}{g^3}.
\end{equation}
This equation can be integrated into
\begin{equation}
G^2 = - \frac{R^4}{g^2} + C_2
\label{gg}\end{equation}
with an integration constant $C_2$. Similarly the string equation 
(\ref{sts}) can be arranged into
\begin{equation}
- \frac{\ddot{f}}{f} = \frac{1}{g^2G} \frac{\pa}{\pa\sigma}(g^2G') 
+ \omega^2,
\label{ff}\end{equation}
which should be also constant to be equated with $C_3$. Therefore we can
choose a solution $f = \cos\sqrt{C_3}\tau$ whose normalization has been
absorbed into the unknown function $g(\sigma)$. The requirement of 
(\ref{ar}) determines the $\tau$-dependence of $t$ so that we have
$t = \tan \sqrt{C_3}\tau$ and 
\begin{equation}
\sqrt{C_3} = \frac{A}{g^2}.
\label{ca}\end{equation}
Combining (\ref{gg}) and (\ref{ff}) we obtain the second order 
differential equation for $G$
\begin{equation}
G''( C_2 - G^2 ) + 2GG'^2 = ( C_2 - G^2 )( C_3 - \omega^2 )G.
\label{seo}\end{equation}
In the constraint (\ref{tt}) $\dot{t}$ is eliminated by (\ref{ar}) with
(\ref{ca}) to give
\begin{equation}
\frac{g^2}{R^2}\left[ - \frac{C_3}{f^2} + \left( \frac{\dot{f}G}{f}
\right)^2 \right] + R^2\left(\frac{\dot{f}}{f}\right)^2 =  
- \frac{g^2}{R^2}[G'^2 + \omega^2G^2 ] - R^2\left(\frac{g'}{g}\right)^2.
\label{ss}\end{equation}
Through (\ref{gg}) this equation can be expressed in terms of $G$ and 
$f$. The requirement that the $\tau$-dependent terms should be canceled
out in the left hand side of it fixes $C_2$ as $C_2 = 1$. The resulting
$\sigma$-dependent first order differential equation becomes 
\begin{equation}
G'^2 + ( \omega^2 + C_3 )G^2 - \omega^2G^4 - C_3 = 0.
\label{gw}\end{equation}
Since differential of (\ref{gw}) with repect to $\sigma$ and (\ref{gw})
itself combine to yield 
(\ref{seo}) with $C_2 = 1$, it suffices to solve the equation 
(\ref{gw}). There remains a task to manipulate the string equation
(\ref{ste}), which is similarly expressed as 
\begin{eqnarray}
R^2\left[ f\frac{\pa}{\pa\tau}\left( \frac{\dot{f}}{f^2}\right) -
g\frac{\pa}{\pa\sigma}\left( \frac{g'}{g^2} \right) + 
\left(\frac{\dot{f}}{f}\right)^2 - \left(\frac{g'}{g}\right)^2 \right] 
\nonumber \\
+ \frac{g^2}{R^2}\left[ \frac{C_3}{f^2} - \left(\frac{\dot{f}}{f}
\right)^2G^2 + G'^2 - \omega^2G^2 \right] = 0.
\label{stn}\end{eqnarray}
Summing the $\tau$-dependent terms including $f$ we notice the 
disappearance of the $\tau$-dependent factors so that (\ref{stn}) also 
reduces to (\ref{seo}). Therefore in order 
to solve the ordinary differential equation (\ref{gw}), 
we transform it through $G(\sigma) = \sqrt{C_3}
G_0(u)/\omega$ with $u = \omega\sigma$ into
\begin{equation}
\left(\frac{dG_0}{du}\right)^2 = ( 1 - G_0^2 )\left( 1 - 
\frac{C_3}{\omega^2} G_0^2 \right).
\end{equation}
In view of this equation the solution can be expressed in terms of 
Jacobi's elliptic function as 
\begin{equation}
G_0 = \sn(\omega\sigma,k), \hspace{1cm} k = \frac{\sqrt{C_3}}{\omega}.
\end{equation}

Here gathering together and putting $\sqrt{C_3} = \ka$ we can express
the soliton solution as 
\begin{eqnarray}
t &=& \tan \ka\tau, \hspace{1cm} l = \frac{\ka\sn(\omega\sigma)}
{\omega\cos \ka\tau}, \nonumber \\
r &=& \frac{R^2\cos\ka\tau}{\dn(\omega\sigma)}.
\label{sol}\end{eqnarray}
Since the action (\ref{ac}) is expressed as $I = \int dtd\sigma
L_0,\; L_0 = \cos^2\ka\tau L/\ka$ where $\tau$ is regarded as a function
of $t$, the canonical momenta conjugate to $r, \phi, l$ are given by
\begin{equation}
\Pi_r = \frac{\pa L_0}{\pa(\pa_tr)} = \frac{R^2\dot{r}}{2\pi\alpha'r^2},
\; \Pi_{\phi} = \frac{r^2l^2\dot{\phi}}{2\pi\alpha'R^2}, \;
\Pi_l = \frac{r^2\dot{l}}{2\pi\alpha'R^2}.
\end{equation}
The conserved spacetime energy of the soliton is given by
\begin{equation}
E = \frac{\cos^2\ka\tau}{4\pi\alpha'\ka}\int_0^{2\pi}
 d\sigma\left[ \frac{r^2}{R^2}
(\dot{t}^2 + l^2\dot{\phi}^2 ) + \frac{R^2}{r^2}( \dot{r}^2 + r'^2 )
+ \frac{r^2}{R^2}( \dot{l}^2 + l'^2 ) \right],
\end{equation}
which turns out, through the constraint (\ref{tt}) and the solution 
(\ref{sol}), to be
\begin{equation}
E = \frac{R^2\ka}{2\pi\alpha'}\int_0^{2\pi}d\sigma \frac{1}
{(\dn(\omega\sigma))^2}.
\label{ene}\end{equation}
The spin of the soliton is also obtained by
\begin{equation}
S = \frac{R^2\omega}{2\pi\alpha'}\int_0^{2\pi}d\sigma 
\left(\frac{k\sn(\omega\sigma)}{\dn(\omega\sigma)}\right)^2.
\label{spi}\end{equation}
Owing to $0 < \dn \le 1$ it is possible to introduce a function 
$\rho(\sigma)$ as $\cosh \rho(\sigma) = 1/\dn(\omega\sigma) =
1/\sqrt{1-k^2\sn^2(\omega\sigma)}$ with $k = \ka/\omega < 1$ and rewrite
(\ref{ene}), (\ref{spi}) as
\begin{eqnarray}
E &=& \frac{R^2\ka}{2\pi\alpha'}\int_0^{2\pi}d\sigma \cosh^2\rho,
\nonumber \\
S &=& \frac{R^2\omega}{2\pi\alpha'}\int_0^{2\pi}d\sigma 
\sinh^2\rho.
\end{eqnarray}
Thus we have recovered the expressions for the energy and spin of the
soliton that were constructed in the global coordinates \cite{FT},
where $\rho$ is the radial coordinate.

The periodicity condition $\rho(\sigma + \pi) = \rho(\sigma)$ for the 
folded closed string that is divided into four segments, combines with
the fundamental periodicity $2K$ of $\dn(\omega\sigma,k)$ where $K$ is the
complete elliptic integral of the first kind, to yield a relation
$\omega\pi = 2K$ which indeed agrees with the periodicity condition 
$\ka = (1/\sqrt{\eta})F(1/2,1/2,1,-1/\eta), \; 1 + \eta = 1/k^2$ 
presented in Ref. \cite{FT}. Similarly from (\ref{sol}) $r$ and $l$ also
satisfy the periodicity conditions, $r(\sigma + \pi) = r(\sigma)$
and $l(\sigma + 2\pi) = l(\sigma)$. The Jacobi elliptic function 
$\dn(\omega\sigma)$ starts at $\sigma=0$ with the maximal value,
$\dn(0)=1$ and gradually decreases to reach the first minimum at
$\sigma = K/\omega$ that is equal to $\pi/2$. Due to $\dn(K) = 
\sqrt{1-k^2}$, the maximal radial coordinate $\rho_0 = \rho(\pi/2)$
is specified by $\cosh\rho_0 = 1/\sqrt{1 - k^2}$. Short strings 
correspond to $k \ll 1$, while long strings are achieved by taking $k$ 
to be the ciritical value 1. The expression $\sinh \rho = 
k\sn(\omega\sigma)/\sqrt{1-k^2\sn^2(\omega\sigma)}$ provides 
$\rho \approx  (1/\sqrt{\eta})\sin\sigma$ for short strings.

Now we consider a closed string rotating along the $\var$-cycle of $S^5$
and stretched along the angular coordinate $\theta$ of $S^5$.
We start to make an ansatz for a time-dependent embedding
\begin{eqnarray}
t &=& t(\tau), \; r = r(\tau,\sigma), \; \theta = \theta(\sigma),
\nonumber \\
\var &=& \nu \tau, \; x_i = \var_i = 0 \; (i=1,2,3)
\end{eqnarray}
with a constant angular velocity parameter $\nu$. From one 
energy-momentum constraint $T_{\tau\sigma} = 0$, 
that is $\dot{r}r'R^2/r^2 = 0$ we can take the 
radial coordinate to be a function of $\tau$, $r = r(\tau)$.
Then the relevant action is expressed as
\begin{equation}
I = - \frac{1}{4\pi\alpha'} \int d\tau d\sigma \left[
\frac{r^2}{R^2}\dot{t}^2 - R^2\sin^2\theta \dot{\var}^2 - 
\frac{R^2}{r^2}\dot{r}^2 + R^2\theta'^2 \right],
\end{equation}
which does not depend on $t$ and $\var$ so that $r^2\dot{t} =
A$ should be $\tau$-independent and $\dot{\var}\sin^2\theta$ is indeed
$\tau$-independent. The other 
constraint $T_{\tau\tau} + T_{\sigma\sigma} = 0$ is given by
\begin{equation}
-\frac{r^2}{R^2} \dot{t}^2 + R^2(\dot{\var}^2\sin^2\theta + \theta'^2 )
 + \frac{R^2}{r^2}\dot{r}^2 = 0.
\label{em}\end{equation}
The $\theta$ equation of motion $\theta'' = - \nu^2\sin\theta\cos\theta$
has a first integral $\theta'^2 = -\nu^2\sin^2\theta + C_1$ with an
integration constant $C_1$. On the other hand the $r$ equation of motion
is given by
\begin{equation}
\frac{r}{R^2}\dot{t}^2 + R^2\left[ \pa_{\tau}\left(\frac{\dot{r}}{r^2}
\right) + \frac{\dot{r}^2}{r^3} \right] = 0,
\label{req}\end{equation}
whose first integral is $\dot{r}^2 = A^2/R^4 - (r/C_2)^2$ with an 
integration constant $C_2$. The further integration yields 
$r = (AC_2/R^2)\cos((\tau + \tau_0)/C_2)$ with an integration constant
$\tau_0$. Substituting these solutions into (\ref{em}) we note that
the constraint is satisfied only when $C_1 = 1/C_2^2$.
If we choose $C_2 = 1/\ka, A = R^4\ka, 
\tau_0 =0$, then the soliton solution
reads $t = \tan\ka\tau, r = R^2\cos\ka\tau$, which is compared with
(\ref{sol}). The canonical momenta conjugate to $r, \var$ are also given
by $\Pi_r = \dot{r}R^2/(2\pi\alpha'r^2), \Pi_{\var} = \dot{\var}
R^2\sin^2\theta/2\pi\alpha'$ which provide the energy and angular momentum
of this system as $E = \ka R^2/\alpha', J = (\nu R^2/2\pi\alpha')
\int d\sigma\sin^2\theta$ accompanied with $\theta'^2 = - 
\nu^2\sin^2\theta + \ka^2$. These expressions are the same as those 
presented using the global metric in Ref. \cite{SIA}.

\section{Orbiting strings in the near-horizon brane \\ backgrounds}

Using similar procedures we consider the closed string motion in the
near-horizon geometry of NS5-branes whose metric is expressed in terms of
the Schwarzschild-type coordinates as
\begin{eqnarray}
ds^2 &=& -dt^2 + \sum_{i=1}^{5}dx_i^2 + Q_5^{NS}\frac{dr^2}{r^2} +
Q_5^{NS}d\Omega_3^2, \nonumber \\
d\Omega_3^2 &=& \sin^2\theta d\var^2 + d\theta^2 + \cos^2\theta d\psi^2
\end{eqnarray}
with the charge of NS5-brane $Q_5^{NS}$. The classical string 
configuration specified by $ t = t(\tau),\; x_i = 0 \;(i=1,\cdots,5),
\;r = r(\tau,\sigma), \;\var = \nu\tau, \;
\theta = \pi/2, \;\psi = 0$ extends along the radial $r$ direction and
rotates along the $\var$ angle with constant angular velocity $\nu$.
The relevant action is given by
\begin{equation}
I = -\frac{1}{4\pi\alpha'}\int d\tau d\sigma \left[ \dot{t}^2 -
Q_5^{NS}\sin^2\theta\dot{\var}^2 + \frac{Q_5^{NS}}{r^2}
( -\dot{r}^2 + r'^2 ) \right],
\end{equation}
where our configuration does not couple to the NS-NS B-field.
Because of one conformal constraint given by $\dot{r}r'Q_5^{NS}/r^2 = 0$,
$r$ cannot have both $\tau$ and $\sigma$ dependences, so that
we choose $r = r(\sigma)$. Since the Lagrangian does not
depend on $t$ and $\var$, it follows that $\dot{t}$ and $\dot{\var}$
are $\tau$-independent. Therefore $t$ can be parametrized by 
$t = \ka\tau$ in the same way as $\var = \nu\tau$.
The $r$ equation of motion is given by
\begin{equation}
\pa_{\sigma}\left(\frac{r'}{r^2}\right) + \frac{r'^2}{r^3} = 0,
\label{rr}\end{equation}
that is compared with (\ref{req}), while the other conformal constraint
is written by
\begin{equation}
r'^2 = \frac{\ka^2 - \nu^2Q_5^{NS}}{Q_5^{NS}}r^2,
\label{cc}\end{equation}
whose differential with respect to $\sigma$ yields (\ref{rr}).
For $\ka^2 \ge \nu^2Q_5^{NS}$ the differential equation (\ref{cc}) has two
exponential solutions $r = r_0\exp(\pm\Omega\sigma)$ with
$\Omega =((\ka^2 - \nu^2Q_5^{NS})/Q_5^{NS})^{1/2}$. In Ref. \cite{KM}
the planetoid string solution for the 
spherical Rindler spacetime showed the
similar exponential behaviors and a folded open string configuration
was constructed. Here splitting the interval $0 \le \sigma \le 2\pi$
into two segments and combining the two solutions we construct a
folded closed string configuration as follows:
\begin{equation}
r(\sigma) = \left\{ \begin{array}{rl}   r_0e^{\Omega\sigma}, & 
\mbox{for $0 \le \sigma \le \pi$}, \\
r_0e^{\Omega(2\pi-\sigma)},
 & \mbox{for $\pi \le \sigma \le 2\pi$}. 
\end{array} \right.
\label{fcs}\end{equation}
The string is stretched from $r_0$ to $r_0e^{\Omega\pi}$ and
doubles back on itself. Thus for $r_0 > 0$
the folded closed string orbits 
outside the origin. The energy and angular momentum (spin) 
of this string configuration are obtained by $E = \ka/\alpha', 
J = \nu Q_5^{NS}/\alpha'$ where each segment yields the same 
contribution. It is convenient to define a length of the
radial range as $l = r_0(e^{\Omega\pi} - 1)$.
 
Gathering together these expressions we derive a dispersion relation
\begin{equation}
E = \sqrt{\frac{Q_5^{NS}}{(\pi\alpha')^2}\ln^2\left( 
\frac{l + r_0}{r_0}\right) + \frac{J^2}{Q_5^{NS}} }.
\label{dis}\end{equation}
Due to the Virasoro constraint the induced metric on the worldsheet of
the string is given by $ds^2 = (\ka^2-\nu^2 Q_5^{NS})
(-d\tau^2 + d\sigma^2)$.
From the string length element $dl_{phy} = \sqrt{\ka^2 - \nu^2Q_5^{NS}}
d\sigma$ we have the physical string length $\Omega\sqrt{Q_5^{NS}}
\pi$ that is compared with $l$.
The long and short strings are specified by $\Omega \gg 1$ and
$\Omega \ll 1$ respectively. The point-like string characterized by
$\Omega = 0$ orbits at the radial location $r = r_0$ and has an
energy-spin relation $E = J/\sqrt{Q_5^{NS}}$. Recently the same
energy-spin relation has been presented in Ref. \cite{MA} where further
the quantum fluctuations around this point-like classical solution
have been studied and shown to reproduce the spectrum of closed strings
in the pp-wave background which was obtained by taking the Penrose 
limit along a particular class of null geodesics at a 
fixed radius on the NS5-brane background. The string theory in the 
obtained pp-wave background is dual to a subsector of LST theory.

Let us turn to the investigation of a closed string motion in the
near-horizon geometry of D$p$-branes whose metric is written by
\begin{eqnarray}
ds^2 &=& \frac{r^{\frac{7-p}{2}}}{\sqrt{Q_p}}
( - dt^2 + \sum_{i=i}^{p}dx_i^2 )
+ \frac{\sqrt{Q_p}}{r^{\frac{7-p}{2}}}( dr^2 + r^2d\Omega_{8-p}^2 ),
\nonumber \\
d\Omega_{8-p}^2 &=& \sin^2\theta d\var^2 + d\theta^2 +
\cos^2\theta d\Omega_{6-p}^2, \label{met} \\
d\Omega_{6-p}^2 &=& \sum_{i=1}^{6-p}\prod_{j=1}^{i-1}\cos^2\var_j 
d\var_i^2 \nonumber
\end{eqnarray}
with the charge of D$p$-brane $Q_p$. We will consider a closed string
configuration characterized by
\begin{eqnarray}
t &=& \ka\tau, \; r = r(\sigma),  \; \var = \nu\tau,\; \theta = 
\frac{\pi}{2}, \nonumber \\
x_i &=& 0 \; (i=1,\cdots,p), \; \var_i = 0 \; (i=1,\cdots,6-p).
\label{con}\end{eqnarray}
The variation with repect to $\var$ of the relevant action
\begin{equation}
I = - \frac{1}{4\pi\alpha'}\int d\tau d\sigma \left[ 
\frac{r^{\frac{7-p}{2}}}{\sqrt{Q_p}}\dot{t}^2 + \frac{\sqrt{Q_p}}
{r^{\frac{7-p}{2}}}( r'^2 - r^2\dot{\var}^2 ) \right]
\end{equation}
yields $\pa_{\tau}(\dot{\var}r^{(p-3)/2}) = 0$, which 
indeed holds for $r = r(\sigma)$ with $p \ne 3$, while it is 
possible for $r$ to have the
$\tau$-dependence in the special $p=3$ case as argued above.
The Virasoro constraint reads 
\begin{equation}
r'^2 = \frac{\ka^2}{Q_p}r^{7-p} - \nu^2r^2,
\label{vc}\end{equation}
whose differential with respect to $\sigma$ combines with (\ref{vc}) 
to give the $r$ equation of motion
\begin{equation}
\frac{\ka^2}{\sqrt{Q_p}}\frac{7-p}{2} r^{\frac{5-p}{2}} - \sqrt{Q_p}
\left[ 2\frac{\pa}{\pa\sigma}\left(\frac{r'}{r^{\frac{7-p}{2}}}\right)
+ \frac{7-p}{2}\frac{r'^2}{r^{\frac{9-p}{2}}} + \frac{p-3}{2}
\frac{\nu^2}{r^{\frac{5-p}{2}}} \right] = 0.
\end{equation}
From $r'^2 \ge 0$ in (\ref{vc}) for $p \le 4$ the radial $r$
location of the string
is restricted to $r \ge r_0 \equiv (Q_p\nu^2/\ka^2)^{1/(5-p)}$.
We can use (\ref{vc}) to find $r$ as a function of $\sigma$
\begin{equation}
\sigma = \pm \frac{\sqrt{Q_p}}{\ka} \int_{r_0}^r \frac{dr'}
{r'\sqrt{r'^{5-p}-r_0^{5-p}}},
\end{equation}
which could be inverted to find
\begin{equation}
r = \frac{r_0}{\left[\cos\left(\pm\frac{\nu(5-p)\sigma}{2}\right)
\right]^{\frac{2}{5-p}} }.
\label{cos}\end{equation}
Combining the two solutions in the two divided segments we construct
a folded closed string orbiting outside the origin as follows:
\begin{equation}
r(\sigma) = \left\{ \begin{array}{rl} 
\frac{r_0}{\left[\cos\frac{\nu(5-p)}{2}\sigma\right]^{\frac{2}{5-p}}},&
\mbox{for $0\le \sigma \le \pi$,}\\
\frac{r_0}{\left[\cos\frac{\nu(5-p)}{2}(2\pi-\sigma)\right]
^{\frac{2}{5-p}}},& \mbox{for $\pi\le \sigma \le 2\pi$.}
\end{array} \right.
\end{equation}
For the D5-brane background the solution of (\ref{vc}) provides the same
closed string configuration as (\ref{fcs}) with $\Omega = 
((\ka^2 - \nu^2Q_5)/Q_5)^{1/2}$ and the integration constant $r_0$.
These solutions are substituted into
the following expressions of the energy and spin on $S^{8-p}$
of the configurations
\begin{eqnarray}
E &=& \frac{\ka}{2\pi\alpha'\sqrt{Q_p}}\int_{0}^{2\pi} d\sigma
r^{\frac{7-p}{2}},\nonumber \\
J &=& \frac{\nu\sqrt{Q_p}}{2\pi\alpha'}\int_{0}^{2\pi} d\sigma
r^{\frac{p-3}{2}}.
\label{eji}\end{eqnarray}

The D5-brane background gives the energy and spin of closed string
\begin{eqnarray}
E &=& \frac{\ka r_0}{\pi\alpha'}\frac{1}{\sqrt{Q_5}\Omega}
(e^{\Omega\pi} - 1), \nonumber \\
J &=& \frac{\nu r_0}{\pi\alpha'}\frac{\sqrt{Q_5}}{\Omega}
(e^{\Omega\pi} - 1),
\label{ejo}\end{eqnarray}
where we have also taken into account of the same contribution from the
two segments of a one-fold string configuration. 
Although the expressions in (\ref{ejo}) are different from those for
the NS5-brane background they yield a similar dispersion relation
\begin{equation}
E = \sqrt{\left(\frac{l}{\pi\alpha'}\right)^2 + \frac{J^2}{Q_5}},
\end{equation}
which is compared with (\ref{dis}). It again reduces to a linear relation
$E \simeq J/\sqrt{Q_5}$ for the point-like or short strings.

For the near-horizon geometry of D4-branes  the
classical closed string is characterized by the following 
energy and spin 
\begin{eqnarray}
E &=& \frac{\nu}{2\ka}J + \frac{r_0}{\pi\alpha'}
\frac{\sin\frac{\nu}{2}\pi}{\cos^2\frac{\nu}{2}\pi}, \nonumber\\
J &=& \frac{Q_4\nu}{\ka\pi\alpha'}\ln \left(
\frac{1 + \sin\frac{\nu}{2}\pi}{ 1 - \sin\frac{\nu}{2}\pi}\right)
\label{dfo}\end{eqnarray}
with $r_0 = Q_4\nu^2/\ka^2$, where we have 
performed the integral for half of 
string multiplied by factor 2. When $\nu \rightarrow 0$ the closed
string shrinks to a point. For short strings that are characterized
by $\nu \ll 1, E$ and $J$ are expressed as $E \approx \nu J/\ka, 
J \approx Q_4\nu^2/\ka\alpha'$. They give an energy-spin relation
$\sqrt{Q_4} E \approx (\alpha'/\ka)^{1/2}J^{3/2}$. Long strings are 
achieved by taking $\nu$ to be the critical value 1. We parametrize
the limit as $\nu = 1 - \eta, \eta \rightarrow 0$. The spin of long closed
string is large as $J \approx (2Q_4/\ka\pi\alpha')\ln(1/\pi\eta)$ and
the energy is also large to be described in terms of $J$ as
\begin{equation}
E \approx \frac{1}{2\ka}J + \frac{4Q_4}{\ka^2\pi\alpha'}
e^{\frac{\ka\pi\alpha'}{Q_4}J}.
\label{lon}\end{equation}
Thus we note that both $E$ and $J$ diverge for long strings in the same 
way as both the energy and the spin on $AdS$ diverge for long strings
in the $AdS_5\times S^5$ background \cite{SIA,FT}.

The D3-background also gives the $\sigma$-dependent solution 
$r = r_0/\cos(\pm\nu\sigma)$ with $r_0 = 
\sqrt{Q_3}\nu/\ka$ of (\ref{cos}) besides the $\tau$-dependent 
solution of (\ref{req}), $r = R^2\cos\ka\tau$.
The $\sigma$-dependent solution is compared with the semi-infinite
open string solution orbiting outside the horizon at $r = l$ in the
2+1 dimensional BTZ black hole \cite{BTZ} with mass $M_{BTZ} = 1$ and
angular momentum $J_{BTZ} = 0$, that is expressed in terms of Jacobi's
elliptic functions as $r = l\dn(\bar{\sigma})/k\cn(\bar{\sigma})$
with $\bar{\sigma} = \pm\ka\sigma/l$ and $k^2 = (\ka^2 - \nu^2l^2)/\ka^2$
in Ref. \cite{KM}. The integrations in (\ref{eji}) with $p = 3$
are simply performed to give
\begin{equation}
E = \frac{\sqrt{Q_3}\nu}{\ka\pi\alpha'}\tan \nu\pi, \hspace{1cm}
J = \frac{\sqrt{Q_3}\nu}{\alpha'},
\end{equation}
that yield a general energy-spin relation in a closed form
\begin{equation}
E = \frac{J}{\ka\pi}\tan \left( \frac{\pi\alpha'}{\sqrt{Q_3}}J
\right).
\end{equation}
Since the induced metric on the worldsheet  is described by
$ds^2 = \nu^2\sqrt{Q_3}\tan^2\nu\sigma(-d\tau^2 + d\sigma^2)$
for $0 \le \sigma \le \pi$, the physical string length is obtained by
$l_{phy} = Q_3^{1/4}\ln (1/\cos\nu\pi)$. For short strings,
$\nu \ll 1$ it reduces to $\sqrt{Q_3}E \approx (\alpha'/\ka)J^2$.
Long strings specified by $\nu \rightarrow 1/2$ have very large
energy and finite spin. 

For the D2-brane background the energy and spin of the closed 
string configuration are evaluated as
\begin{eqnarray}
E &=& - \frac{\nu}{2\ka}J + \frac{r_0}{\pi\alpha'}
\frac{\sin\frac{3}{2}\nu\pi}{(\cos\frac{3}{2}\nu\pi)^{2/3}},
\nonumber \\
J &=& \frac{2\ka r_0}{3\pi\alpha'\nu}\sqrt{2\alpha}F_1
\left(\frac{1}{2},-\frac{1}{3},\frac{1}{2},\frac{3}{2};
\alpha,\frac{\alpha}{2}\right),
\end{eqnarray}
where $r_0^3 = Q_2\nu^2/\ka^2, \alpha = 1-\cos (3\nu\pi/2)$
and $F_1$ is Appell's hypergeometric function.  
From them we extract an energy-spin relation $\sqrt{Q_2}E 
\approx (\alpha'/\ka)^{3/2}J^{5/2}$
for short strings, $\nu \ll 1$. For long strings specified by
$\nu \rightarrow 1/3$ the energy diverges, 
while the spin remains finite.

The D1-brane background also gives 
\begin{eqnarray}
E &=& - \frac{\nu}{\ka}J + \frac{r_0}{\pi\alpha'}\frac{\sin 2\nu\pi}
{\sqrt{\cos 2\nu\pi}}, \nonumber \\
J &=& \frac{\sqrt{Q_1}}{\pi\alpha'r_0}\left[ \sqrt{2}E(\alpha,
\frac{1}{\sqrt{2}}) - \frac{1}{\sqrt{2}}F(\alpha,\frac{1}{\sqrt{2}})
\right]
\end{eqnarray}
with $r_0^4 = Q_1\nu^2/\ka^2$, where $F$ and $E$ are the elliptic integral
of the first and second kinds respectively, and $\sin\alpha = \sqrt{2}
\sin\nu\pi$. Short strings have a similar energy-spin relation 
$\sqrt{Q_1}E \approx (\alpha'/\ka)^2J^3$.
Long strings obtained by $\nu \rightarrow 1/4$ have very large
energy with finite spin.

Gathering together we observe that the energy-spin relations for short
orbiting strings in the near-horizon geometries of D$p$-branes
($p = 1,\cdots, 5$) can be summarized in a compact form as
\begin{equation}
\sqrt{Q_p}E \approx \left(\frac{\alpha'}{\ka}\right)^{\frac{5-p}{2}}
J^{\frac{7-p}{2}},
\end{equation}
where $p=5$ is the special case such that the energy is linearly related
to the spin. The intriguing linear behavior $\sqrt{Q_5}E \approx J$
for the point-like string orbiting at the radial location $r=r_0$
in the D5-brane background can be directly compared with the
spectrum of closed strings in the geometry produced by taking the
Penrose limit along a special null geodesic which stays at a fixed 
radius. This kind of special Penrose limit yields a solvable string
theory and we write down here the string spectrum presented 
in Ref. \cite{OS}
\begin{equation}
\sqrt{s}E - J = \sum_n \left[ e^{U_0} \frac{s}{J}N_n^{r,\omega}|n| +
\sqrt{1+e^{2U_0}\frac{s^2}{J^2}n^2}N_n^y \right],
\end{equation}
where $s$ is the number of D5-branes and $U_0$ specifies the
fixed radial coordinate, and $N_n^{r,\omega}$ as well as $N_n^y$
are the oscillation occupation numbers of massless and massive 
scalars respectively. Thus the ground state satisfying the classical
relation $\sqrt{s}E = J$ corresponds to the classical point-like 
string configuration.

Moreover, we consider a closed string motion in the near-horizon
geometry of NS1-branes with metric
\begin{equation}
ds^2 = \frac{r^6}{Q_1^{NS}}(-dt^2 + dx^2) + dr^2 + r^2(\sin^2\theta
d\var^2 + d\theta^2 + \cos^2\theta d\Omega_5^2),
\end{equation}
where $d\Omega_5^2$ is described in terms of $\var_i \; (i=1,\cdots,5)$
by an expression similar to $d\Omega_{6-p}^2$ in (\ref{met}) and
$Q_1^{NS}$ is the charge of NS1-brane. For the closed string 
configuration similarly expressed by (\ref{con}) with $p=1$, 
which does not couple to the NS-NS B-field, we see
that the radial equation of motion essentially
becomes the same as that for the 
D1-brane background, although the two metrics are different.
The energy and spin of this configuration are given by
\begin{eqnarray}
E &=& \frac{\nu}{2\ka}J + 
\frac{\sqrt{Q_1^{NS}}\nu^2}{4\ka^2\pi\alpha'}
\frac{\sin 2\nu\pi}{\cos^2 2\nu\pi}, \nonumber \\
J &=& \frac{\sqrt{Q_1^{NS}}\nu}{4\ka\pi\alpha'}\ln 
\frac{1+\sin 2\nu\pi}{1-\sin 2\nu\pi},
\end{eqnarray}
which have structures similar to (\ref{dfo}) for the D4-brane 
background. For short strings they give an energy-spin relation
$(Q_1^{NS})^{1/4}E \approx (\alpha'/\ka)^{1/2}J^{3/2}$.
Long string are produced by taking a limit $\nu \rightarrow 1/4$
and have an energy-spin relation 
\begin{equation}
E \approx \frac{1}{8\ka}J + \frac{\sqrt{Q_1^{NS}}}{(16\ka)^2\pi\alpha'}
e^{\frac{16\ka\pi\alpha'}{\sqrt{Q_1^{NS}}} J},
\end{equation}
which shows the same behavior as (\ref{lon}).

\section{Conclusion}

Manipulating the Schwarzschild-type coordinates for the 
$AdS_5\times S^5$ background, we have analyzed the motion of 
the closed string rotating on $AdS_5$ and stretched simultaneously
on the radial coordinate $l$ in the non-compact directions of the 
D3-branes and the radial coordinate $r$ in the transversal directions.
We have demonstrated that such two radial coordinates 
for the soliton solution are characterized by products of the
trigonometrical function of the worldsheet time $\tau$ and Jacobi's 
elliptic function of the worldsheet coordinate $\sigma$, while
in the global coordinates the corresponding radial coordinate $\rho$
is explicitly described by Jacobi's elliptic function of $\sigma$.
This product form is a new type of behavior which is not seen
in the planetoid string solutions in various curved spacetimes 
\cite{VLS,VE,KM}.
In the Schwarzschild-type coordinates we have also reconstructed a 
classical solution of the closed string rotating on $S^5$ and 
stretched along an angular direction, whose radial coordinate $r$ is
specified by the trigonometrical function of $\tau$, which corresponds
to the rotating point-like string staying at the origin of the
radial coordinate $\rho$ in the global coordinates.

Based on the similar prescription we have studied the orbiting closed
strings on the near-horizon geometries of D$p$-branes $(1\le p \le 5)$,
NS1-branes and NS5-branes, which are expressed in terms of the
Schwarzschild-type coordinates, and shown that their radial behaviors
are characterized by the trigonometrical function of $\sigma$ for
D$p$-branes $(1\le p \le 4)$ and NS1-branes, and the exponential
function of $\sigma$ for D5-branes and NS5-branes.
We have noted that the D3-brane background is so special that there
are two type of closed string solutions, a $\tau$-dependent configuration
for the radial coordinate and a $\sigma$-dependent one.
There is some resemblance between the energy-spin relation for the 
D4-brane background and that for the NS1-brane background.
Although the classical energies of closed strings are implicitly given
by the complicated functions of the spins, we have extracted
a simple general formula of energy-spin relation for short strings
in the D$p$-brane backgrounds. The energies of short strings show 
power behaviors of the spins and specially the D5-brane as well as 
NS5-brane backgrounds yield a linear behavior for point-like 
strings. We have observed that this linear relation coincides
with the linear one for the ground state energy of the closed string
spectrum in the pp-wave backgrounds produced by taking a special Penrose
limit along the null geodesics such that the radial coordinate is 
constant on the near-horizon geometries of the D5-branes and NS5-branes.
It is compared with the linear relation seen in the energy and angular
momentum  of the point-like string sitting at $\rho = 0$ and rotating
on $S^5$ for the $AdS_5\times S^5$ background in the global coordinates
\cite{SIA}.

\end{document}